\documentclass[conference]{IEEEtran}
\IEEEoverridecommandlockouts
\usepackage{cite}
\usepackage{amsmath,amssymb,amsfonts}
\usepackage{algorithmic}
\usepackage{graphicx}
\usepackage{textcomp}
\IEEEpubid{\makebox[\columnwidth]{978-1-5386-3779-1/18/\$31.00~\copyright{}2018 IEEE \hfill} \hspace{\columnsep}\makebox[\columnwidth]{ }}
\def\BibTeX{{\rm B\kern-.05em{\sc i\kern-.025em b}\kern-.08em
    T\kern-.1667em\lower.7ex\hbox{E}\kern-.125emX}}

\title{Multirate 5G Downlink Performance Comparison for f-OFDM and w-OFDM Schemes with Different Numerologies
\thanks{Marina Mondin is also affiliated with DET, Politecnico di Torino, C.so Duca degli Abruzzi 24, 10129, Torino, Italy (marina.mondin@polito.it).}
}

\author{\IEEEauthorblockN{Francesco Di Stasio}
\IEEEauthorblockA{\textit{DET,Politecnico di Torino} \\
\textit{C.so Duca degli Abruzzi 24}\\
10129 Torino, Italy\\
francesco.distasio@polito.it}
\and
\IEEEauthorblockN{Marina Mondin }
\IEEEauthorblockA{\textit{ECE, California State Univ. Los Angeles} \\
\textit{5151 State University Dr. }\\
Los Angeles, 90032, CA, US\\
marina.mondin@calstatela.edu}
\and
\IEEEauthorblockN{Fred Daneshgaran}
\IEEEauthorblockA{\textit{ECE, California State Univ. Los Angeles} \\
\textit{5151 State University Dr. }\\
Los Angeles, 90032, CA, US\\
fdanesh@calstatela.edu}
}

\begin{document}
\maketitle

\begin{abstract}
One of the main open problems for next generation wireless networks, is to find the new OFDM-based waveform to be used in 5G. The new modulation scheme must primarily be able to achieve higher spectral efficiency than its predecessor. The main 3GPP's candidate is a new version of OFDM, called Filtered Orthogonal Frequency-Division Multiplexing (f-OFDM), which is similar to OFDM but with additional filtering in order to reduce Out-Of-Band (OOB) emissions and to obtain a better spectral-localization. Another option is windowed-OFDM (w-OFDM), which is basically a classical OFDM scheme where each symbol is windowed and overlapped in the time domain. In this paper we compare classic OFDM signals using Cyclic Prefix (CP-OFDM) with f-OFDM and w-OFDM, each one with multiple parametric options and numerologies. A multirate transmitter simultaneously operating with multiple numerologies is considered, where the transmitted sub-bands must be up-sampled and interpolated in order to generate the composite numerical signal fed to the Digital to Analog Converter (DAC). Finally, we discuss advantages and disadvantages of the various schemes.
\end{abstract}


\section{Introduction}
Next generation cellular networks present extremely challenging issues for researchers and engineers. The main goal is to improve the actual LTE performance, in order to meet the growing data demand from the newly provisioned technologies and services. For example, increasing the data rate by a factor 100 with respect to LTE, while decreasing the latency from the actual 15 ms down to as low as approximately, 1 ms. Requirements and key features of 5G are fully presented in \cite{whatwill5Gbe}, \cite{5Groadmap}, \cite{fiveDisruptive} and \cite{5GkeyEnabling}.
The new services foreseen in 5G and their associated requirements can be summarized as follows:
\begin{itemize}
\item \textit{URLLC (Ultra Reliable Low Latency Communications)}: requires low latency and high reliability;
\item \textit{eMBB (Enhanced Mobile Broadband)}: requires low latency, high spectral efficiency and high data rate;
\item \textit{mMTC (massive Machine Type Communication)}: requires low energy consumption, low device complexity and an improved link budget.
\end{itemize}

Enabling new technologies, such as massive MIMO, Device-to-Device communications (D2D), Wireless Software Defined Networking (WSDN), Millimeter Wave communications and network Densification, are being utilized in order to reach 5G's goals. \\
In this paper we deal with problems regarding Radio Access techniques. As stated earlier, new services in 5G require high data rates and large spectral efficiency. For this reason, we focus on spectral efficiency problem of a legacy OFDM (\textit{Orthogonal Frequency-Division Multiplexing}) system, which has to improve its performance to achieve the required targets. As is well known, OFDM is one of the most important transmission techniques of the recent past \cite{OFDMforWireless}, largely used in LTE standards. The principle of OFDM based on sub-carrier division has been well studied and tested during the years and the first advantage of this scheme is its simplicity of implementation. Moreover, OFDM allows for simple modulation and demodulation and is MIMO friendly. On the other hand, OFDM suffers from high PAPR (Peak-to-Average Power Ratio) and, mostly, of high Out-Of-Band (OOB) emissions. The required Cyclic Prefix (CP) and strict bounds for synchronization are among other disadvantages of OFDM. Indeed, in a 5G scenario it is desirable to use sub-bands that do not need to be perfectly synchronized with each other due to the different requirements of the multitude of devices or services on the network. In fact, in 5G we will have different kind of devices which rarely connect to the network. For example, an IoT (Internet of Things) device needs to send just few control bytes on rare occasions, and several kind of devices will have very short battery life. For these reasons, it may be desirable to use a waveform with relaxed synchronization requirements.
\\
This article attempts to summarize benefits and disadvantages of the two schemes currently being considered by 3GPP (\textit{Third Generation Partnership Project}) for 5G applications, namely f-OFDM (\textit{Filtered OFDM}) and w-OFDM (\textit{Windowed OFDM}), and to compare them with classic CP-OFDM. In the last case, we consider standard OFDM sub-bands, without using any strategy to reduce OOB emissions. 
In the f-OFDM schemes we use low-pass filters in order to attenuate the OOB emissions and have an efficient sub-band divided system, as explained in \cite{zhang2015filtered}, \cite{abdoli2015filtered} and \cite{RBbasedfOFDM}, while, in the w-OFDM scenario, OOB emissions are reduced by smoothing the symbol transitions  with a time domain window applied on each sub-band, according to \cite{UTwindowing} and \cite{tdWindowingIEEE}. Other results on f-OFDM can be found in \cite{zhangfiltered} which gives a closed form for ISI (Inter-Symbol Interference), ICI (Inter-Carrier Interference) and ACI (Adjacent-Channel Interference). \cite{FOFDMbank} suggests a filter-bank version of f-OFDM, while \cite{PAPRredFOFDM} discusses PAPR reduction in f-OFDM. Finally, \cite{FOFDMfield} and \cite{5GFieldTrials}, describe field trials results for f-OFDM and w-OFDM based systems in real environments. 
\\
In contrast to previous studies, our comparison is made for a realistic multirate transmitter able to simultaneously transmit sub-bands with different numerologies, where the signals associated with the low-rate sub-bands must be up-sampled and interpolated in order to generate the composite numerical signal fed to the DAC. As a consequence, our results also account for the effects generated by up-sampling and interpolation.  
The remainder of this paper is organized as follows. In section \ref{sec:TX} we present our framework and show the generic structure of the considered Base Station (BS) transmitter. In subsections \ref{sec:CPOFDM}, \ref{sec:FOFDM} and \ref{sec:WOFDM}, we give details on the considered signal waveforms, and in the section \ref{sec:RX} we describe the receiver structure. Finally, in section \ref{sec:numerical} we show different simulation results to evaluate and compare our waveforms, and finally present the conclusions in section \ref{sec:conc}.

\section{Transmitter structure}\label{sec:TX}

\begin{figure*}[t]
\centering
\includegraphics[scale=0.35]{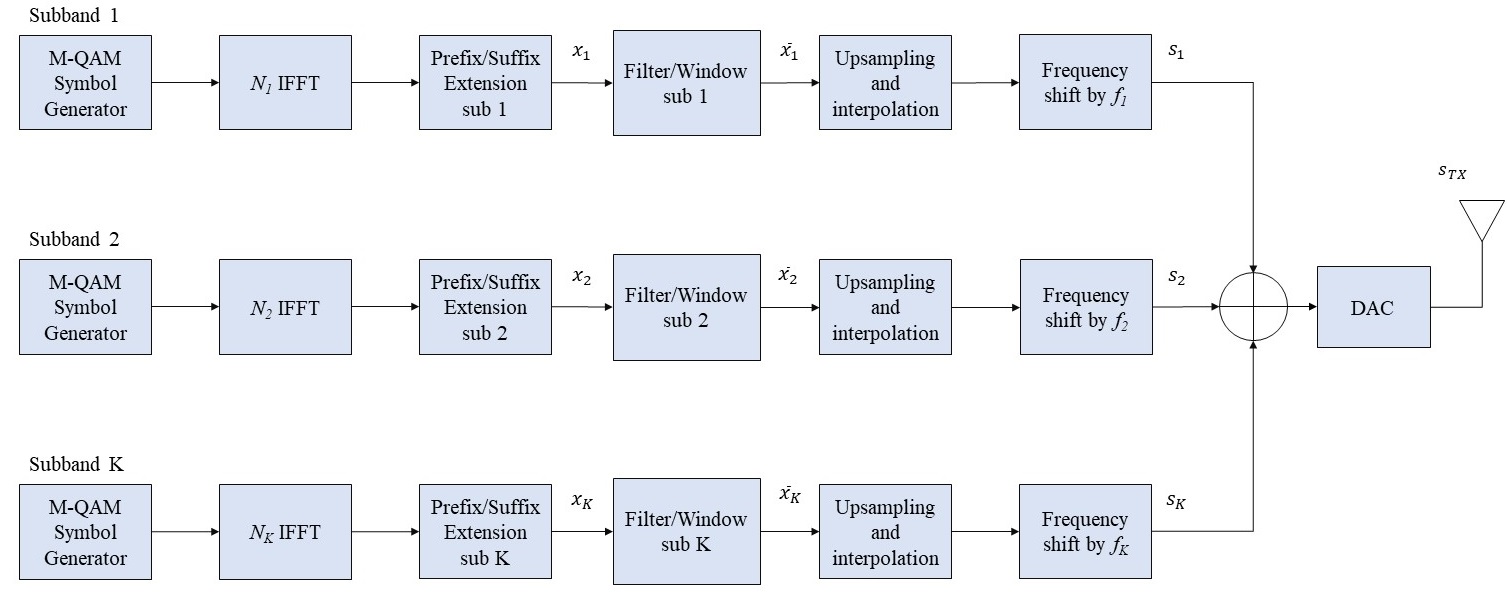}
\caption{Transmitter block diagram using frequency-localized OFDM-based waveforms.}
\label{fig:genTX}
\end{figure*}

In this section we provide a full description of our three proposed schemes, we start by showing the common elements and then detail the specifics of each signal. On the transmitter, we are focusing on a Base Station (BS) scenario, which simultaneously transmits multiple sub-bands with multiple numerologies, followed by a single-sub-band receiver (we are assuming that each service, or user, could be served by more than one sub-band). \\
Figure \ref{fig:genTX} shows the BS transmission structure which is common for all the considered waveforms, with differences residing in the \textit{prefix/suffix} and \textit{filter/window} blocks. The processing details for the considered schemes are described in sections \ref{sec:CPOFDM}, \ref{sec:FOFDM} and \ref{sec:WOFDM}.
\\
The transmission steps are very similar to a classical OFDM scheme, starting with a QAM symbol generator over the $K$ branches, where $K$ represents the number of sub-bands generated by the BS and it depends on the plurality of services offered by the network. In order to have a fair comparison, we assume to have the same modulation order $M$ among all the sub-bands. We will consider M-QAM schemes, with $4 \leq M \leq 256$, as suggested by the 3GPP LTE-A standard. These symbols are transmitted on the useful sub-carriers (and possibly on the pilot sub-carriers, that in this study will be treated as useful sub-carriers), while null values are formally transmitted on the (unused) null sub-carriers. \\
Since we will have different coexisting numerologies, every sub-band may have an FFT/IFFT processor with a different dimension, so we denote by $N_i$ the IFFT dimension used in the $i^{th}$ sub-band. $N_i$ is typically selected to be a power of 2, so we can write $N_i= 2^{q_i}$, where $q_i$ is a non-negative integer. Each sub-band could have a different number $N_{u_i}$ of useful sub-carriers, so that we have $(N_i-N_{u_i})$ null sub-carriers. Furthermore, $N_{c_i}$ represents the total number of `'guard'' sub-carriers introduced between the $i^{th}$ sub-band and its neighboring sub-bands. We choose to use $N_{c_i}/2$ sub-carriers on the right and the same number on the left side of our useful sub-carriers. We also denote as $\Delta f_i$ the sub-carrier spacing in the $i^{th}$ sub-band, defined as a power-of-2 multiple of the minimum sub-carrier spacing $f_0$ (equal to $15kHz$), so that we can write $\Delta f_i = f_{0} 2^{p_i}$, where $p_i$ is a non-negative integer. Given these assumptions, we can express the sampling frequency $f_{s_i}$ used for the $i^{th}$ sub-band as:

\begin{equation}
\label{eq:sampfreq}
f_{s_i}=N_i \cdot \Delta f_i = f_{0} \cdot 2^{p_i + q_i}.
\end{equation}

Given these assumptions, the transition guard band between the $i^{th}$ and $(i+1)^{th}$ sub-band (shown in Figure 2) is equal to $ \Delta_{i(i+1)}=(N_{c_i}+N_{c(i+1)})/2$. 

We also denote as $W$ the total bandwidth used for transmission, and as $W_i$ the total bandwidth of the $i^{th}$ sub-band. With this representation we have $W_i = \Delta f_i \cdot (N_{u_i}+ N_{c_i}) $ and $\sum _{i=1}^{K} Wi = W$. Obviously, in order to achieve a better spectral efficiency it is desirable to have  $N_{c_i} = 0  \; \forall i$ in order to avoid wasting frequency resources, with a consequential loss in terms of spectral efficiency. In this paper we consider different values of $N_{c_i}$ to show their impact on some performance indicator such as the raw-BER.
\\
The \textit{prefix/suffix} block adds a Cyclic Prefix or Suffix depending on the considered waveform. 
Next, with reference to Figure \ref{fig:genTX}, the sequence $\mathbf{x_i}$ is either filtered or windowed depending on the considered scheme in order to improve frequency localization and lower the spectral side-lobes. 
\\
In order to add the signals associated to sub-bands with different numerologies we have to convert them to a common sampling frequency (possibly the least common multiple frequency) that we achieve by up-sampling the signals sampled at the lower frequencies.
Given our assumptions, the least common multiple frequency is the maximum frequency among all the sub-bands, i.e., from (\ref{eq:sampfreq})

\begin{equation}
\label{eq:fsmax}
f_s^{\max}=\max_{1 \leq i \leq K} \left( f_{s_i} \right) = f_0  2^{ \left( p_m + q_m \right)} \,,
\end{equation} 
where, $m=\arg \max_i \left( p_i + q_i \right) $. The upsampling factor for each sub-band is calculated as

\begin{equation}
\label{eq:Ups}
U_{s_i}=2^{\left(\left( p_m + q_m \right)- 
\left( p_i + q_i \right)\right)}
\end{equation}

Each up-sampled signal needs to be interpolated  with a low pass filter (designed as described in \ref{sec:FOFDM}) and properly shifted in the frequency domain by frequency $f_i$ in order to generate the composite overall digital sequence, where 




\begin{equation}
\label{eq:f_i}
f_i=f_{i-1} + \Delta f_{i-1} \frac{N_{u_{i-1}} + N_{c_{i-1}}}{2}
+ \Delta f_{i} \frac{N_{u_i} + N_{c_i}}{2} \,,
\end{equation}
and $f_1$, is the center frequency of the first sub-band. The signals associated to the $K$ sub-bands are then added and sent to the DAC. 


\subsection{CP-OFDM} \label{sec:CPOFDM}

In the case denoted as CP-OFDM, the transmitter implements a classic OFDM scheme with cyclic prefix (without windowing or filtering to reduce OOB emissions). We will consider this scheme as reference against which we compare the other schemes. In this case, the \textit{Prefix/Suffix} block just adds the Cycling Prefix (CP) containing $N_{g_i}$ samples.

\subsection{Filtered OFDM} \label{sec:FOFDM}

The transmission chain for f-OFDM is very similar to that for the CP-OFDM, with an additional low-pass filter introduced after the CP concatenation and before the frequency shift in order to reduce the OOB emissions. Clearly, the structure of the transmitter low-pass filter is very important for reducing OOB emissions and possible interference. Ideally, we want a filter perfectly flat in pass-band and zero outside this band, with null transition bands. This kind of filter is physically unrealizable, but can be approximated by truncating and windowing the ideal $sinc(\cdot)$ impulse response.
\\
This operation introduces a new element in this framework, the filter transition bands. It is important to note that the transition bands are completely independent from frequency guard bands introduced above. Obviously having the transition band contained in the guard band could guarantee better performances.
\\
The filter has to be as flat as possible in the pass-band with tight transition bands. To achieve this goal we have chosen a windowed-sinc filter with ideal impulse response
\begin{figure}[h]
\centering
\label{fig:transitionband}
\includegraphics[scale=0.36]{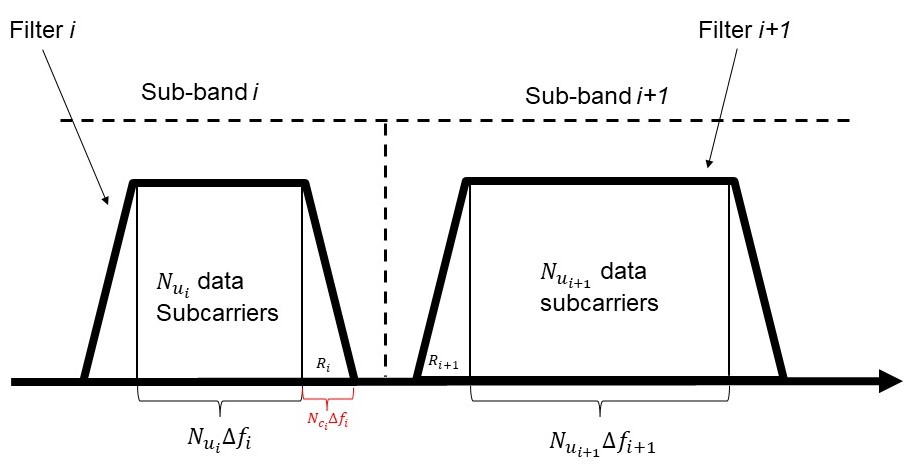}
\caption{Role of transition bands of the filter.}
\end{figure}
\begin{equation}
\label{eq:filter1}
p_i(n)= sinc\left(\Delta f_i \left(N_{u_i} + 2R_i\right) \frac{n}{N_i} \right)\;  ,
\end{equation}
for $-\lfloor L_i/2 \rfloor \leq n \leq \lfloor L_i/2 \rfloor$
where, $L_i$ represents the filter order and $\Delta f_i R_i$ the transition band on one side. $p_i(n)$ does not represent our final filter, it is just a truncated $sinc$. The final coefficients of our normalized lowpass filter are given by
\begin{equation}
\label{eq:filter2}
f_i(n)=\frac{p_i(n)\cdot w_i(n)}{\sum_k p_i(k)\cdot w_i(k)}\,,
\end{equation}
where, the term $w_i(n)$ represents the chosen window. In this paper we use a raised cosine window defined as
\begin{equation}
\label{eq:FilterWindow}
w_i(n)=\left( 0.5 \left( 1 + \cos \left( 
\frac{2\pi n}{L_i - 1} \right) \right) \right)^{0.6},
\end{equation}
where, $n$ is bounded as in equation (\ref{eq:filter1}).
The filter impulse response contains $2L_i + 1$ samples, which causes a signal extension in the time domain by $2L_i$ samples. This signal elongation could seem a critical point of the f-OFDM scheme leading to interference in time. Fortunately, this kind of filter has the major part of its energy concentrated in the main Sinc lobe, so the elongation is important just for a small time period during the CP of the successive symbol (see Figure \ref{fig:simLength}). For this reason, it is not strictly necessary to choose $L_i$ to be very small, specifically $L_i$ can be larger than $N_{g_i}$ (length of the cyclic prefix).


\subsection{Windowed OFDM} \label{sec:WOFDM}

\begin{figure}[t]
\centering
\includegraphics[scale=0.25]{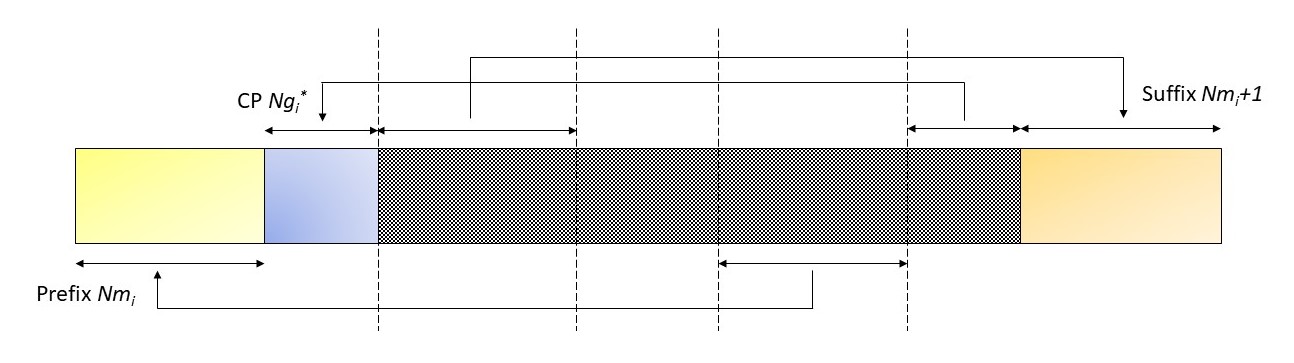}
\caption{CP, prefix and suffix extension for a w-OFDM symbol.}
\label{fig:prefixSuffix}
\end{figure}

In this section we illustrate the time domain windowing strategy. Since, the signal high frequency components are mainly generated by the discontinuities between adjacent OFDM symbols, softening these singularities with a smooth transition lowers the OOB emissions. Specifically, the OFDM symbols must be elongated with the insertion of CP, prefix and suffix, then windowed and finally concatenated (by partially overlapping two consecutive symbols).
\\
The first operation is to extend our OFDM symbol by copying the last $N_{g_i}$ samples of the native OFDM symbol at the beginning of the new w-OFDM symbol, as typically done for CP-OFDM (\ref{sec:CPOFDM}). The first $N_{m_i}$ samples are denoted as ''prefix'', while the remaining $N_{g_i}$ are denoted as CP. The w-OFDM symbol is then further extended by copying the first $N_m+1$ samples of the native OFDM symbol at the end of the new w-OFDM symbol, as shown in Figure \ref{fig:prefixSuffix}. 

Native OFDM symbols in each sub-band may have different lengths, hence the parameter $N_{m_i}$ is used to denote the prefix/suffix parameter for the $i^{th}$ sub-band.
At this point the w-OFDM symbol that we denote as $\mathbf{x}_i$ contains $N_i^{(W)}=N_i + N_{g_i} + 2N_{m_i} + 1$ samples. However, prefix and suffix will be smoothed with a windowing operation, and then the suffix of the $i^{th}$ w-OFDM symbol will be overlapped with the first $N_{m_i} + 1$ samples of the $(i+1)^{th}$ w-OFDM symbol.

The windowed symbol $\mathbf{\bar{x}_i}$ is obtained from the extended symbol $\mathbf{x}$ via 
\begin{equation}
\label{eq:windowing}
\mathbf{\bar{x}_i}=\mathbf{w_i} \cdot
\mathbf{x_i} \,,
\end{equation}
where, $\mathbf{w_i}$ represents the chosen window of length $N_i^{(W)}$. As proposed in \cite{UTwindowing}, we use a window defined via,
\begin{equation}
\label{eq:window}
\mathbf{w_i}= \left[
\begin{array}{c}
\mathbf{0}^{(N_{m_i}-N_{tr_i}/2)}\\
\mathbf{\widetilde{w}}_{tr_i}  \\
\mathbf{1}^{(N_i + N_{g_i} -N_{tr_i} +1)}\\
\mathbf{\widetilde{w}}_{tr_i}^I  \\
\mathbf{0}^{(N_{m_i}-N_{tr_i}/2)}
\end{array}
\right]
\end{equation}
where, $\mathbf{0^L}$ represents a column vector of $L$ elements filled by zeros, likewise $\mathbf{1^L}$ is the same kind of vector filled by ones. The parameter $N_{tr}$ represents the window transition length, i.e. the number of samples the window spends to go from zero-to-one and from one-to-zero, and $N_{tr_i}$ is the transition length in the $i^{th}$sub-band. The vector $\mathbf{\widetilde{w}}_{tr}$ contains the uphill samples of the window, and $\mathbf{\widetilde{w}}_{tr}^I$  the downhill samples. These vectors are defined via,
\begin{equation}
\label{eq:wn1}
\mathbf{\widetilde{w}}_{tr_i}= \left[ 
w_{(i,0)},w_{(i,1)},\dots,w_{(i,N_{tr_i}-1)}
\right]^T
\end{equation}
\begin{equation}
\label{eq:wn2}
\mathbf{\widetilde{w}}_{tr_i}^I= \left[ 
w_{(i,N_{tr_i}-1)},\dots,w_{(i,1)},w_{(i,0)}
\right]^T
\end{equation}

\begin{figure}[h]
\centering
\includegraphics[scale=0.25]{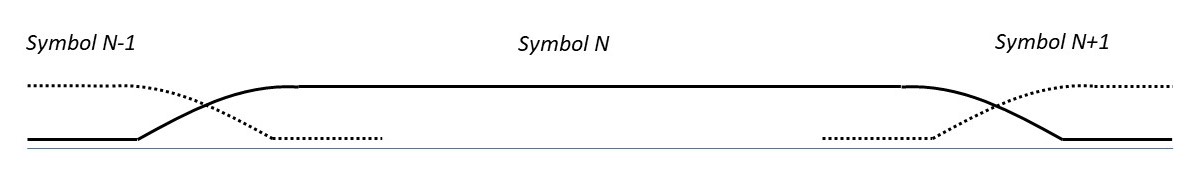}
\caption{Overlapping between adjacent w-OFDM symbols.}
\label{fig:wofdmOverlap}
\end{figure}

\begin{figure}[h]
\centering
\includegraphics[scale=0.4]{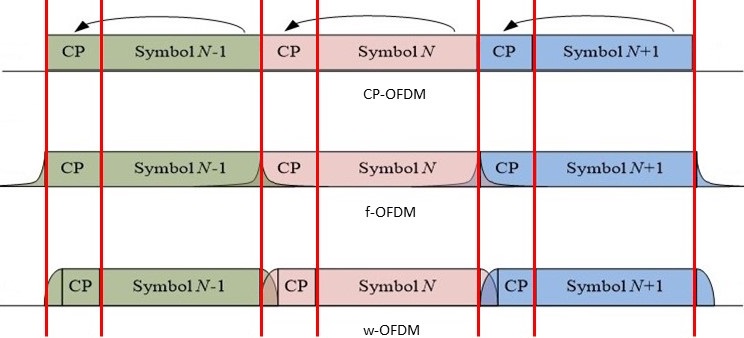}
\caption{Symbols length over different modulations.}
\label{fig:simLength}
\end{figure}

In this paper we chose to use a Blackman windowing function so that,
\begin{equation}
\label{eq:wn3}
w_{(n,i)}=0.42-0.5
\cos \left( \frac{\pi n}{N_{tr_i}}\right)+
0.08\cos \left( \frac{2 \pi n}{N_{tr_i}}\right).
\end{equation}
This completes the explanation of how to build a single w-OFDM symbol.
The last $N_{m_i}+1$ of a symbol are then overlapped and added to the first $N_{m_i}+1$ of the successor, as shown in Figure \ref{fig:wofdmOverlap}. Hence, to obtain the desirable effect of smooth transitions, we must choose the overlapping length $N_{m_i}+1$ to be higher than the transition time $N_{tr_i}$. Is important to notice that the w-OFDM symbol is longer than CP-OFDM and f-OFDM symbols. So that, in order to have comparable symbol duration, for our simulations we decide to adopt a different CP duration for w-OFDM, defined as $Ng_i^* = Ng_i - Nm_i$, with $Nm_i<Ng_i$. In this configuration the final w-OFDM duration will be $N_i+Ng_i+Nm_i+1$, where the last $Nm_i+1$ samples are overlapped with the following symbol. Figure \ref{fig:simLength} shows symbol duration on our three framework.

\section{Receiver structure} \label{sec:RX}
\begin{figure}[h]
\centering
\includegraphics[scale=0.22]{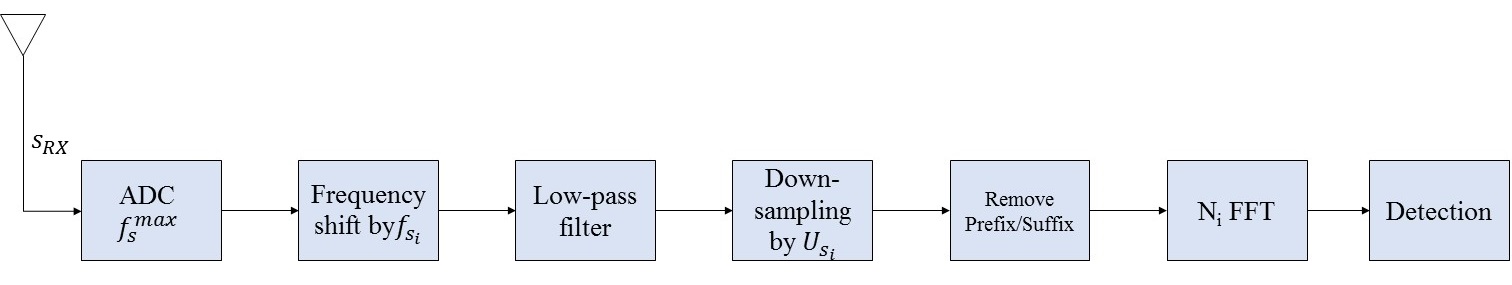}
\caption{Receiver block diagram for the $i^{th}$ sub-band using frequency-localized OFDM-based waveforms.}
\label{fig:genRX}
\end{figure}
Figure \ref{fig:genRX} shows the generic structure of the receiver for a the $i^{th}$ sub-band. The receiver structure is very general and can be adapted to various schemes with minimal modifications. \\
The first step is the frequency down-conversion,
that generates the baseband version of the signal from which to filter out the $i^{th}$ sub-band. After filtering, we need to down-sample,  eliminate the prefix/suffix and perform a $N_i$-points FFT. 

The low pass filter has to be as flat as possible in our pass-band, with tight transition bands and high stopband attenuation. In this paper we choose to use the same filter used in transmission for the f-OFDM in \ref{sec:FOFDM}.

\section{Numerical results} \label{sec:numerical}

In this section we provide the results obtained by numerical simulations of CP-OFDM, f-OFDM and w-OFDM schemes in a three sub-bands environment. Since we are only comparing the robustness of the considered schemes to adjacent channel interference generated by the simultaneous use of different numerologies, we consider a simple AWGN channel, without taking into account transmission effects such as non-linearity or fading, that would affect all the waveforms in the same way.
The simulations consider a scenario with $3$ sub-bands with the numerologies listed in Table \ref{tab:diffNum}. In all simulations we chose to use as number of useful data carriers an integer multiple of a \textit{PRB (Physical Resource Block)}, fixed at $12$ sub-carriers. In particular we chose to use $15$ PRB for each sub-band and for each simulation.   

%
%

\begin{figure}[h]
\centering
\includegraphics[width=.4\textwidth]{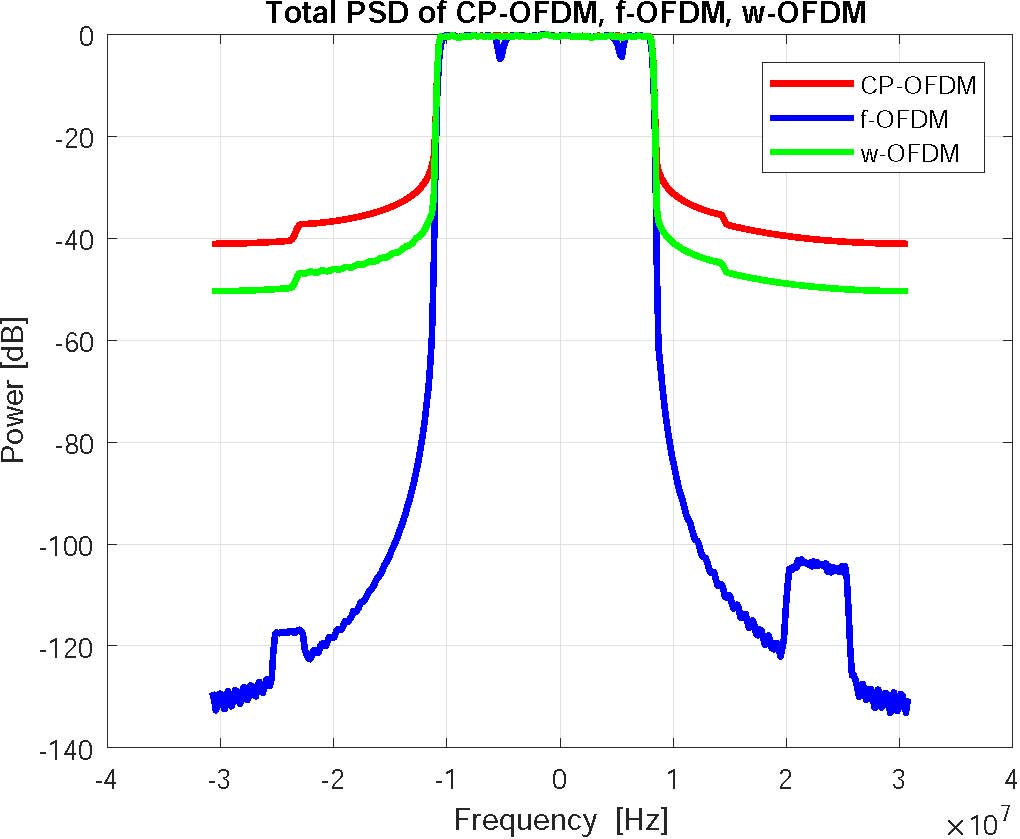}
\caption{PSD of the composite signal with parameter listed in Table 1}
\label{fig:PSD2}
\end{figure}

\begin{figure}[h]
\centering
\includegraphics[width=.4\textwidth]{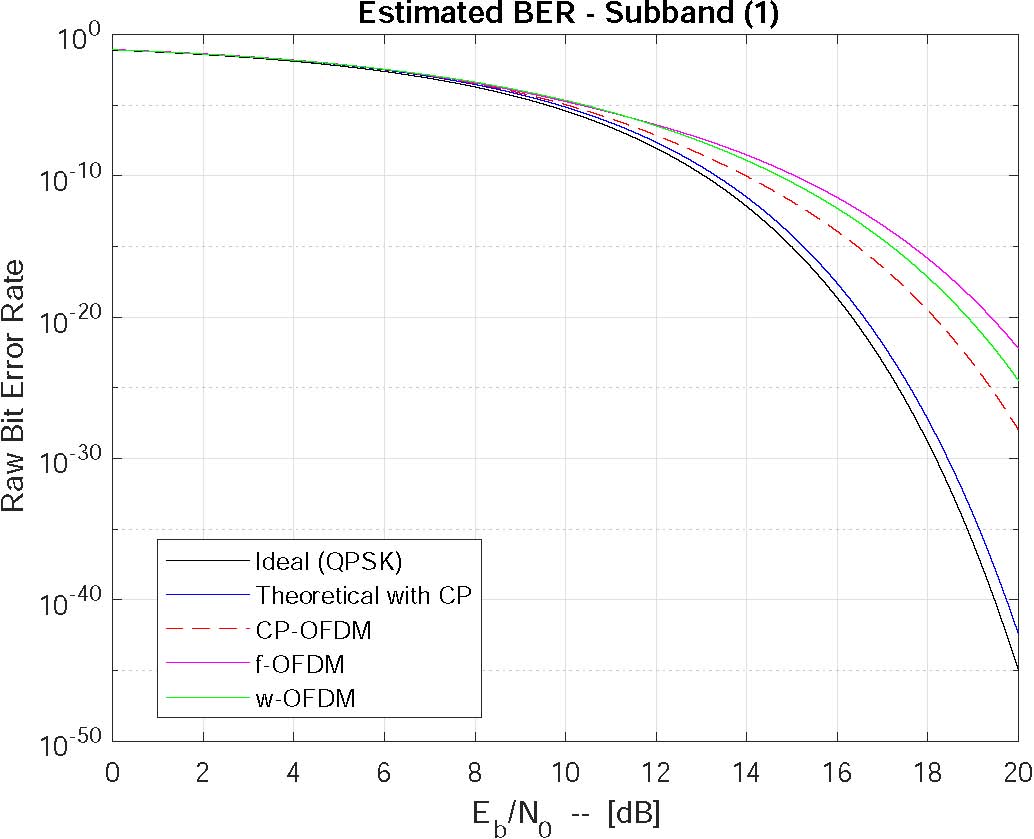}
\caption{BER performances for the subchannels listed in Table 1 (QPSK)}
\label{fig:BER1}
\end{figure}

\begin{figure}[h]
\centering
\includegraphics[width=.45\textwidth]{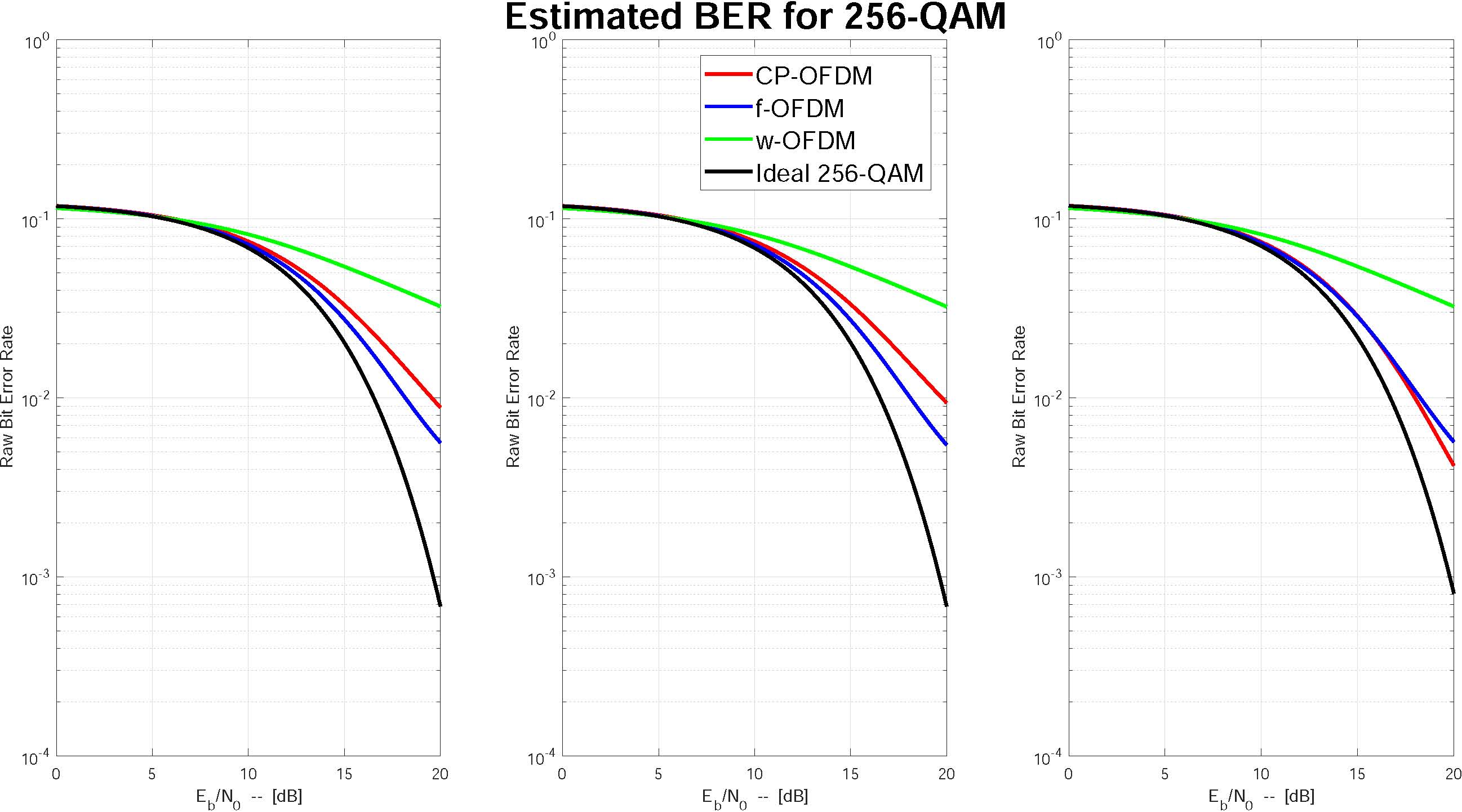}
\caption{BER performances for the subchannels listed in Table 1 (256 QAM)}
\label{fig:BER2}
\end{figure}

The Power Spectral Density (PSD) of the composite signal containing the three subbands described in Table \ref{tab:diffNum} is shown Figure \ref{fig:PSD2}. The bit-error-rate performances obtained using semi-analytic techniques \cite{shan} for QPSK and 256 QAM constellations are reported in Figures \ref{fig:BER1} and \ref{fig:BER2}, respectively.
\begin{table}[h]
\centering
\label{tab:diffNum}
\begin{tabular}{|l||c|c|c|}
\hline
Parameter & Sub-band 1 & Sub-band 2 & Sub-band 3 \\ \hline
$N_i$      & 1024  & 1024  & 1024 \\ \hline
$N_{g_i}$     & 64 & 64  & 64  \\ \hline
$\Delta f_i \,(kHz)$ & 30 & 60 & 15 \\ \hline
$L_i$      & 177 & 89 & 353 \\ \hline 
$\Delta_{i(i-i)}$ \,(kHz) & 180  & 180 & 180 \\ \hline 
$R_{i}$ \,(kHz) & 90  & 90 & 90 \\ \hline 
\end{tabular}
\caption{Numerologies used in the considered simulations.}
\end{table}


As it can be observed in Figures \ref{fig:BER2} and \ref{fig:BER1}, when different numerologies are used in adjacent subchannels (introducing ACI, \textit{Adjacent Channel Interference}, due to the non-orthogonality of the OFDM subcarriers), a $E_b/N_0$ loss can be observed on the BER curves that would not be present with a unique numerology (these ideal curves are not reported for brevity). As expected, such loss is more noticeable with higher modulation orders and in subchannels that suffer from higher interference from adjacent subchannels (like subchannel 3, whose higher ACI can be observed in Figure \ref{fig:BER2}).

\begin{figure}[h]
\centering
\includegraphics[width=.4\textwidth]{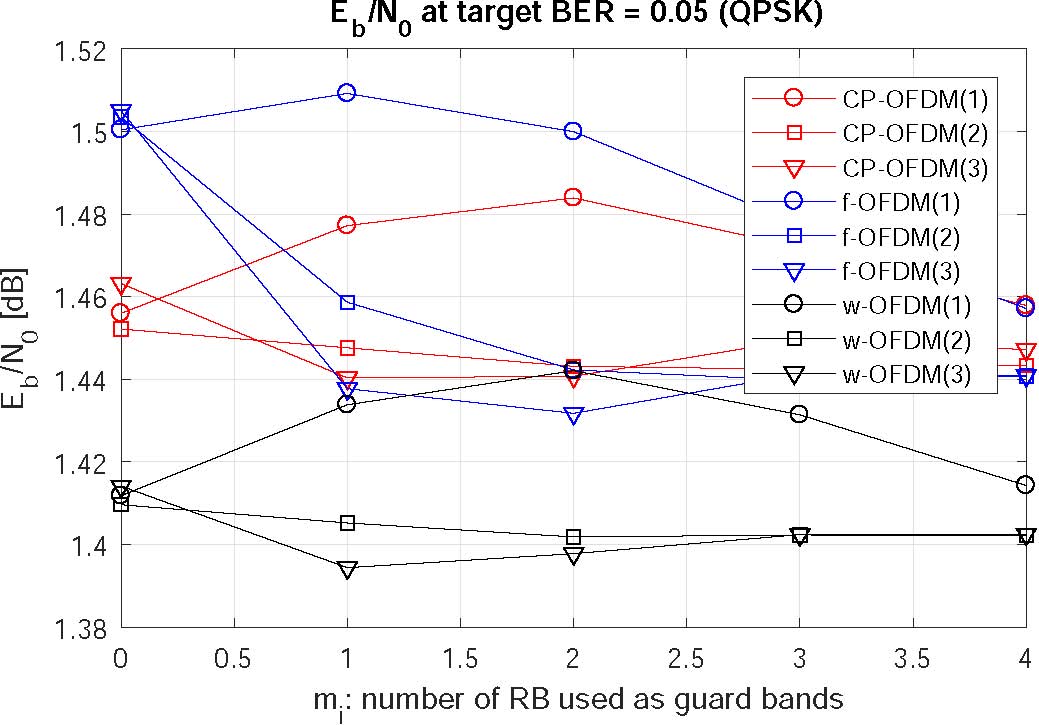}
\caption{$E_b/N_0$ at BER=0.05 s function of $m_i$ (QPSK)}
\label{fig:ebn0QPSK}
\end{figure}

\begin{figure}[h]
	\centering
	\includegraphics[width=.4\textwidth]{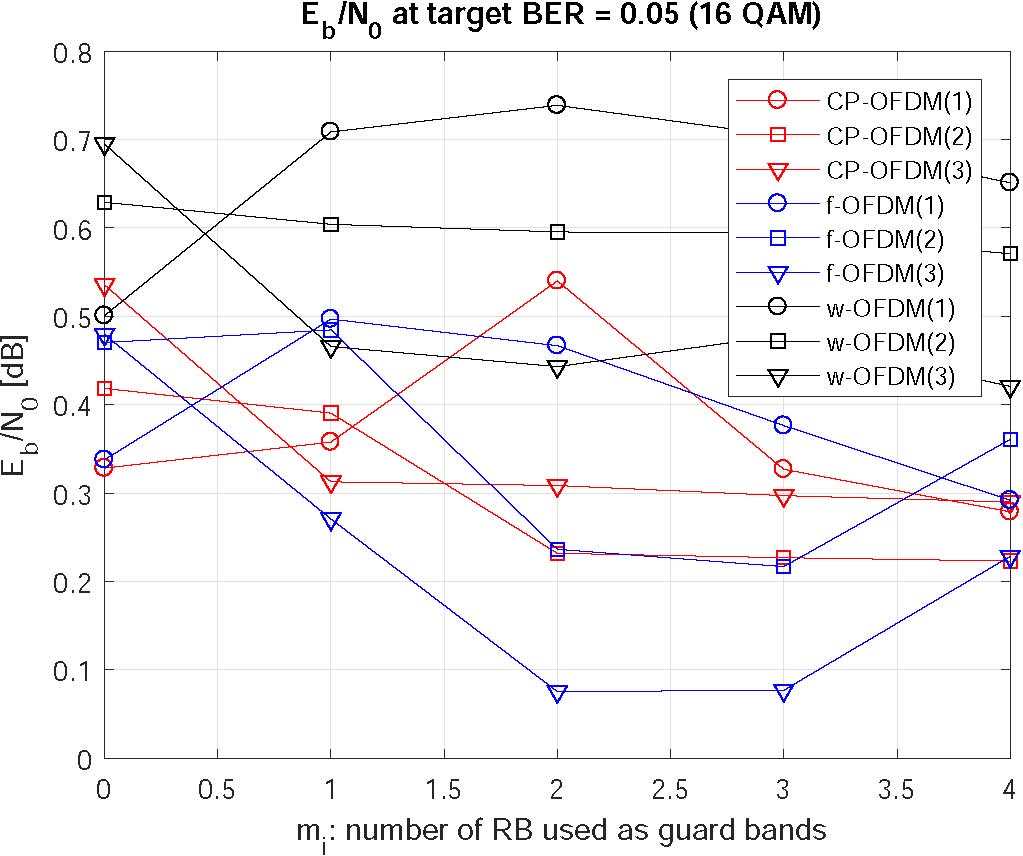}
	\caption{$E_b/N_0$ at BER=0.05 as function of $m_i$ (16 QAM)}
	\label{fig:ebn016QAM}
\end{figure}

\begin{figure}[b]
	\centering
	\includegraphics[width=.4\textwidth]{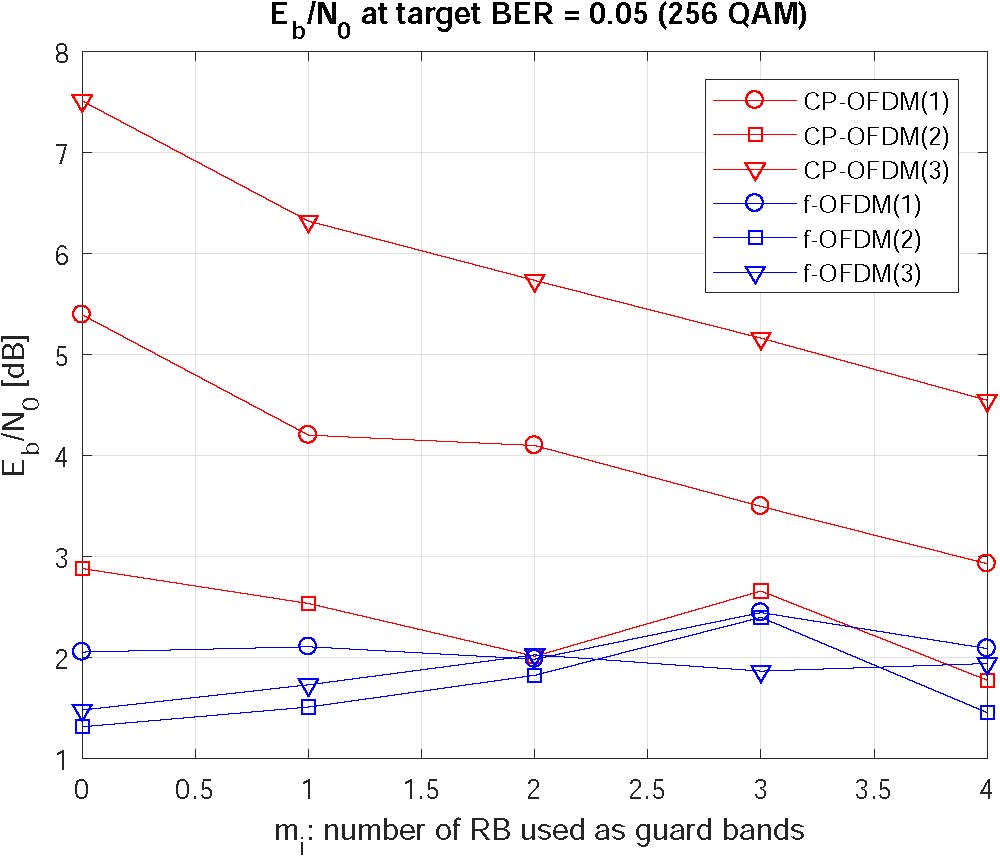}
	\caption{$E_b/N_0$ at BER=0.05 as function of $m_i$ (256 QAM)}
	\label{fig:ebn0256QAM}
\end{figure}

A second set of performance curves has been obtained to show the sensitivity of various schemes to subchannel separation, where we have selected $\Delta_{i(i+1)}=12 m_i f_0$ (i.e., the subchannel separation is equal to $m_i$ resource blocks with $0 \leq m_i \leq 4$) and $R_i=\Delta_{i(i+1)}/2$. \\
As shown in Figure \ref{fig:ebn0QPSK}, BER performance improve with subchannel separation and for low modulation orders (QPSK) w-OFDM offers better performances than f-OFDM (in this case, the distortion introduced by filtering on f-OFDM is more relevant than the distortion introduced on w-OFDM by the use of a windowed CP). The situation changes as the modulation order increases (and the effect of ACI becomes more relevant): as shown in Figure \ref{fig:ebn016QAM}, CP-OFDM offers the best performance in this scenario for 16QAM (while f-OFDM and w-OFDM are almost comparable), while f-OFDM outperforms the other schemes with a 256QAM modulation as shown in \ref{fig:ebn0256QAM} (w-OFDM performance in this case are very bad and are not shown in the graph). Overall we can observe that while w-OFDM outperforms f-OFDM for low modulation orders, f-OFDM offers better performance with high modulation orders. Furthermore, at high modulation orders, the performance of f-OFDM does not significantly degrade as the channel separation decreases, and it is possible to allocate the various subchannels extremely close to each other, thus improving the spectral efficiency. 
\section{Conclusions} \label{sec:conc}
In this paper a multirate transmitter simultaneously operating with multiple numerologies has been considered, comparing the performance of the OFDM-based schemes that are currently being considered for 5G applications in a realistic scenario via accurate simulations. Specifically, simulated performances of CP-OFDM, w-OFDM and f-OFDM schemes have been presented showing that while w-OFDM offers better performance with low modulation orders, f-OFDM, thanks to its better OOB attenuation, outperforms both w-OFDM and CP-OFDM with high modulation orders, while offering a reduced sensitivity to subchannel separation.

\bibliography{references}

\begin{thebibliography}{10}
\providecommand{\url}[1]{#1}
\csname url@samestyle\endcsname
\providecommand{\newblock}{\relax}
\providecommand{\bibinfo}[2]{#2}
\providecommand{\BIBentrySTDinterwordspacing}{\spaceskip=0pt\relax}
\providecommand{\BIBentryALTinterwordstretchfactor}{4}
\providecommand{\BIBentryALTinterwordspacing}{\spaceskip=\fontdimen2\font plus
\BIBentryALTinterwordstretchfactor\fontdimen3\font minus
  \fontdimen4\font\relax}
\providecommand{\BIBforeignlanguage}[2]{{%
\expandafter\ifx\csname l@#1\endcsname\relax
\typeout{** WARNING: IEEEtran.bst: No hyphenation pattern has been}%
\typeout{** loaded for the language `#1'. Using the pattern for}%
\typeout{** the default language instead.}%
\else
\language=\csname l@#1\endcsname
\fi
#2}}
\providecommand{\BIBdecl}{\relax}
\BIBdecl

\bibitem{whatwill5Gbe}
J.~G. Andrews, S.~Buzzi, W.~Choi, S.~V. Hanly, A.~Lozano, A.~C. Soong, and
  J.~C. Zhang, ``{What will 5G be?}'' \emph{{IEEE} Journal on selected areas in
  communications}, vol.~32, no.~6, pp. 1065--1082, 2014.

\bibitem{5Groadmap}
I.~F. Akyildiz, S.~Nie, S.-C. Lin, and M.~Chandrasekaran, ``{5G roadmap: 10 key
  enabling technologies},'' \emph{Computer Networks}, vol. 106, pp. 17--48,
  2016.

\bibitem{fiveDisruptive}
F.~Boccardi, R.~W. Heath, A.~Lozano, T.~L. Marzetta, and P.~Popovski, ``{Five
  disruptive technology directions for 5G},'' \emph{{IEEE} Communications
  Magazine}, vol.~52, no.~2, pp. 74--80, 2014.

\bibitem{5GkeyEnabling}
E.~Hossain and M.~Hasan, ``{5G cellular: key enabling technologies and research
  challenges},'' \emph{{IEEE} Instrumentation \& Measurement Magazine},
  vol.~18, no.~3, pp. 11--21, 2015.

\bibitem{OFDMforWireless}
R.~v. Nee and R.~Prasad, \emph{{OFDM for wireless multimedia
  communications}}.\hskip 1em plus 0.5em minus 0.4em\relax Artech House, Inc.,
  2000.

\bibitem{zhang2015filtered}
X.~Zhang, M.~Jia, L.~Chen, J.~Ma, and J.~Qiu, ``{Filtered-OFDM-enabler for
  flexible waveform in the 5th generation cellular networks},'' in \emph{Global
  Communications Conference (GLOBECOM)}.\hskip 1em plus 0.5em minus 0.4em\relax
  IEEE, 2015, pp. 1--6.

\bibitem{abdoli2015filtered}
J.~Abdoli, M.~Jia, and J.~Ma, ``{Filtered OFDM: A new waveform for future
  wireless systems},'' in \emph{Signal Processing Advances in Wireless
  Communications (SPAWC), 2015 IEEE16th International Workshop on}.\hskip 1em
  plus 0.5em minus 0.4em\relax IEEE, 2015, pp. 66--70.

\bibitem{RBbasedfOFDM}
J.~Li, K.~Kearney, E.~Bala, and R.~Yang, ``{A resource block based filtered
  OFDM scheme and performance comparison},'' in \emph{Telecommunications (ICT),
  2013 20th International Conference on}.\hskip 1em plus 0.5em minus
  0.4em\relax IEEE, 2013, pp. 1--5.

\bibitem{UTwindowing}
K.~Mizutani and H.~Harada, ``{Universal Time-Domain Windowed OFDM},'' in
  \emph{Vehicular Technology Conference (VTC-Fall), 2016 IEEE 84th}.\hskip 1em
  plus 0.5em minus 0.4em\relax IEEE, 2016, pp. 1--5.

\bibitem{tdWindowingIEEE}
K.~Mizutani, Z.~Lan, and H.~Harada, ``{Time-domain windowing design for {IEEE}
  802.11 af based TVWS-WLAN systems to suppress out-of-band emission},''
  \emph{IEICE Transactions on Communications}, vol.~97, no.~4, pp. 875--885,
  2014.

\bibitem{zhangfiltered}
L.~Zhang, A.~Ijaz, P.~Xiao, M.~Molu, and R.~Tafazolli, ``{Filtered OFDM
  Systems, Algorithms and Performance Analysis for 5G and Beyond}.''

\bibitem{FOFDMbank}
Y.~Qiu, Z.~Liu, and D.~Qu, ``{Filtered bank based implementation for filtered
  OFDM},'' in \emph{2017 7th IEEE Int. Conf. on Electronics Information and
  Emergency Communication (ICEIEC)}.\hskip 1em plus 0.5em minus 0.4em\relax
  IEEE, 2017, pp. 15--18.

\bibitem{PAPRredFOFDM}
M.~B. Mabrouk, M.~Chafii, Y.~Louet, and C.~F. Bader, ``{A Precoding-based PAPR
  Reduction Technique for UF-OFDM and Filtered-OFDM Modulations in 5G
  Systems},'' in \emph{European Wireless 2017}, 2017.

\bibitem{FOFDMfield}
D.~Wu, X.~Zhang, J.~Qiu, L.~Gu, Y.~Saito, A.~Benjebbour, and Y.~Kishiyama, ``{A
  field trial of f-OFDM toward 5G},'' in \emph{Globecom Workshops (GC Wkshps),
  2016 IEEE}.\hskip 1em plus 0.5em minus 0.4em\relax IEEE, 2016, pp. 1--6.

\bibitem{5GFieldTrials}
P.~Guan, D.~Wu, T.~Tian, J.~Zhou, X.~Zhang, L.~Gu, A.~Benjebbour, M.~Iwabuchi,
  and Y.~Kishiyama, ``{5G Field Trials: OFDM-Based Waveforms and Mixed
  Numerologies},'' \emph{IEEE Journal on Selected Areas in Communications},
  vol.~35, no.~6, pp. 1234--1243, 2017.

\bibitem{shan}
K.~S.~S. Michel C.~Jeruchim, Philip~Balaban, \emph{{Simulation of Communication
  Systems: Modeling, Methodology and Techniques}}.\hskip 1em plus 0.5em minus
  0.4em\relax Springer US, 2000.

\end{thebibliography}
\bibliographystyle{IEEEtran}


\end{document}